\documentclass[twoside]{amsart}
\usepackage{graphicx}
\usepackage{mathptmx}      
\usepackage{amsmath}
\usepackage{dcolumn}
\usepackage{bm}
\usepackage{bbm}
\usepackage{rotating}
\usepackage{braket}
\usepackage{epsfig}

\title[Polarization-entangled photon generation in a quantum dot-cavity system interacting with laser fields] {Polarization-entangled photon generation by a semiconductor quantum dot coupled to a cavity interacting with external fields}

\author{Kostas Blekos}
\address{Department of Physics, School of Natural Sciences, University of Patras, Patras 265 04, Greece}
\author{Nikos Iliopoulos}
\author{Maria-Eftaksia Stasinou}
\author{Evaggelos Vlachos}
\author{Andreas F. Terzis}

\date{28 December 2013}

\begin{document}

\begin{abstract}
We theoretically  investigate polarization-entangled photon generation by using
a semiconductor quantum dot embedded in a microcavity. The entangled states can
be produced by the application of two cross-circularly polarized laser fields.
The quantum dot nanostructure is considered as a four-level system (ground, two
excitons and bi-exciton states) and the theoretical study relies on the dressed
states scheme. The quantum correlations, reported in terms of the entanglement
of formation, are extensively studied for several values of the important
parameters of the quantum dot system as the bi-exciton binding energy, {the
decoherence times of the characteristic transitions, the quality factor of the
cavity} and the intensities of the applied fields.
\end{abstract}

\maketitle

\section{Introduction}

In the relatively new scientific field of quantum information and quantum
computing \cite {nielsen}, the creation and control of entangled-photon pair is
a rather key issue  \cite{ekert,sleat,cirac,zeilinbook}. The most used (common)
way of generating entangled-photon pairs is through a parametric down
conversion  \cite{zeilin,kwiat}.  An alternative technique for entangled-photon
generation is the two-photon cascade decay of atomic excitations
\cite{kocher,clauser,aspect}. Recently, an extension of the atomic technique,
the cascade-emission process from a biexciton state in a semiconductor quantum
dot (SQD, usually called artificial atoms), was proposed  \cite{benson} and
demonstrated  \cite{akopian,young,hafen,dousse}. In this more efficient
entangled-photon generation scheme, the entangled photons are generated from
the cascade emission through the degenerate intermediate states having
different polarizations.

A much better control of this process is achieved by placing the SQD in a
microcavity, where the material excitations and the photons of the cavity  are
strongly coupled. Then the SQD-cavity system is described in terms of the
coupled states of material excitations and cavity photons, which usually
called dressed states (properly treated using cavity quantum electrodynamics,
i.e. describing matter and light on quantum mechanical grounds). Actually, it
has been shown in recent theoretical investigations that under specific optimal
conditions a drastic enhancement of the entangled-photon generation can be
achieved for SQDs  embedded in a cavity \cite{ajiki2,ajiki3,ajiki4,ajiki5}.
This is achieved, as the dressed-states formed in  the SQD-cavity system
eliminate the which-path information which prevents the formation of
entangled-photon pair in the biexciton decay process. Moreover, the energies of
these intermediate  dressed states are tuned to be degenerate states by using a
high-Q cavity \cite{johne1,johne2,pathak}.

Recently the entangled-photon generation from a SQD-cavity system, where the
biexciton is efficiently excited under resonant conditions, was studied in
Refs. \cite{ajiki2,ajiki3,ajiki4,ajiki5}.  The resonant excitations of
biexciton occur through the dressed
biexciton states.  Two of the four Bell states can be generating by selecting
the frequencies of applied fields with certain polarizations (cross-circularly
polarized fields) \cite{ajiki2,ajiki5}. The other two Bell states can be
generated by changing the combination of polarizations or by properly adjusting
only the frequencies of the input fields\cite{ajiki2,ajiki4}. Furthermore, they
have achieved control of the non-entangled co-polarized photons  \cite{ajiki5},
as they are strongly suppressed due to the photon blockade effect
\cite{imamo,birnb,faraon}. 

In the present paper we investigate theoretically the entangled - photon
generation from an SQD-cavity system by calculating the thermal entanglement of
Formation (EoF). This paper is organized as follows. In Sec. II, we describe
our model of the QD-cavity system and the resulting dressed states. The Bell
states generated from the dressed states are presented in Sec. III. In Sec. IV,
we present numerical results on the entangled-photon pairs from the QD-cavity
system in  the presence of a biexciton. Our results are summarized in Sec.V.

\section{Theory}

We describe the QD nanostructure by a four-level system taking into account the
ground state  $|G\rangle$, the two circularly  right and left polarized
degenerate exciton states  $|X_R\rangle,|X_L\rangle$ and the biexciton state
$|B\rangle$, which are orthogonal and normalized. In addition we consider that
the QD is embedded  in a microcavity which supports circularly right- and left-
polarized photons of energy equal to the energy difference between the exciton
and the ground state. Actually, we should be careful if we have to include
other (excited) exciton states and biexciton states. The present model is valid
once the level separation of excitons and biexcitons are greater than the vacuum
Rabi splitting. These approximations are valid in the GaAs SQDs as for
realistic values of the later dimensions of the SQD the rabi splitting is of
the order of few tenths of meV, while the level separation is a few meV
\cite{gammon}. Then, the QD-cavity system is represented by a Hilbert space
which is the direct product of states of the QD nanostructure and the
cavity-mode photons. Hence the representation of the product states has the
following form,  $|Y,n_R,n_L\rangle$  where Y describes the energy states of
the QD and $n_R(n_L)$ specifies the number of the right- (left-) polarized
photons in the cavity.  In fact the presence of an anisotropic electron-hole
exchange interaction results in the well known fine structure splitting (FSS)
of the two excitonic states. However in our study we assume that the two
exciton states are degenerate, assuming that the FSS has been minimized by
either properly treating the QD structural properties, as for example by
controlling the growth process, or by applying external (mechanical, electric
and magnetic) fields. 
The Hamiltonian describing the QD-cavity quantum system is given by 
\begin{equation}
\begin{split}
H_0 = &  \epsilon_0\sum_k(|X_k\rangle\langle X_k|+\alpha_k^{\dagger} \alpha_k)\\
&+(2 \epsilon_0-\Delta_B)|B\rangle\langle B| \\
&+ g\sum_k(i|G\rangle\langle X_k|\alpha_k^{\dagger} +H.c.) \\
& + g_B (i|X_R\rangle\langle B|\alpha_R^{\dagger} + i|X_L\rangle\langle B|\alpha_L^{\dagger} +H.c.)
\end{split}
\end{equation}
where the index $k$ runs over the $R$ (right- polarized) and $L$
(left-polarized) QD states or cavity-mode photons. $\epsilon_0$ is the energy
difference between the exciton and the ground state. $g$($g_B$) is the coupling
constants between the exciton(biexciton) transition and one cavity-mode photon.
$\Delta_B$ is the binding energy of the biexciton state. The ground state has
zero energy. In this article we adopt the natural units ($\hbar$=1).
It is rather straightforward to diagonalize the $H_0$ Hamiltonian and find its
eigenstates, usually called dressed states. These dressed states being a linear
superposition of product states are characterized and grouped by the total
number of QD excitations and photons. The product states are divided into one
state of zero energy ($|G,0,0\rangle$), four states of $\epsilon_0$ energy
($|G,1,0\rangle$, $|G,0,1\rangle$, $|X_R,0,0\rangle$ and $|X_L,0,0\rangle$),  seven
states of $2\epsilon_0$ energy 
($\ket{G,1,1}$, $|G,2,0\rangle$, $|G,0,2\rangle$, $|X_R,1,0\rangle$, 
$|X_R,0,1\rangle$, $|X_L,1,0\rangle$ and $|X_L,0,1\rangle$) and one state of
$2\epsilon_0-\Delta_B$ energy ($|B,0,0\rangle$). The dressed states are
categorized into three groups. The group with the lowest (zero) energy which is
actually the $|G,0,0\rangle$ product state. The second group with states of
higher energy, all in the region of $\epsilon_0$. Actually we have only two
energy values with a $2g$ energy splitting known as the vacuum Rabi splitting.
These dressed states are a superposition of singly excited QD of specific
polarization and absence of photons and ground state QD in the presence of one
photon with the same polarization and hence can be characterized by their
polarization (namely, ($|R_+\rangle$,$|R_-\rangle$,$|L_+\rangle$, and
$|L_-\rangle$). Finally, we have eight dressed states of energy around
2$\epsilon_0$. We can separate these states into two categories, the
co-polarized dressed states (linear superposition of  $|G,2,0\rangle$,
$|G,0,2\rangle$, $|X_R,1,0\rangle$, and $|X_L,0,1\rangle$) denoted as
$|RR_+\rangle$, $|RR_-\rangle$, $|LL_+\rangle$ and $|LL_-\rangle$ and  the
cross-polarized dressed states (linear superposition of  $|G,1,1\rangle$,
$|B,0,0\rangle$, $|X_R,0,1\rangle$, $|X_L,1,0\rangle$) corresponding to a
singlet state \[|S\rangle=\frac{1}{\sqrt{2}}\left(|X_R,0,1\rangle-|X_L,1,0\rangle\right)\] 
of energy $2\epsilon_0$ and three triplet
states. Additionally, the cross-polarized states have eigen-energies of the form
$\lambda_j=2\epsilon_0-a_j$, where $j=1,2,3$ \cite{ajiki5}. Substantially,
$a_j$ are the differences of $2\epsilon_0$ from each $\lambda_j$ and they are
the real solutions, multiplied by $\Delta_B$, of the equation \cite{ajiki5},
\begin{equation}
f(x)=x^3-x^2-(p+q)x+p=0
\end{equation}
with $p=2(\frac{g}{\Delta_B})^2$ and $q=2(\frac{g_B}{\Delta_B})^2$.

In general the cavity is coupled to the environment by cavity-mode photons
leaking out of the cavity. In order to have a control of these leaking photons
we apply external electromagnetic fields interacting with the microcavity. We
consider CW laser fields with electric fields $E_Re^{-i\Omega_R t}$ and
$E_Le^{-i\Omega_L t}$ , where  $E_R(E_L$) and $\Omega_R(\Omega_L)$  are the
amplitudes and the frequencies of the right-(left-) polarized laser fields all
assumed real. Then the Hamiltonian describing the interaction between the
lasers and the cavity photons is  \cite{ajiki5,ajiki6}
\begin{equation}
	H_{int}= \sqrt{\Gamma}\sum_{k=R.L} E_k(ie^{-i\Omega_k t}\alpha_k^{\dagger} - ie^{i\Omega_k t}\alpha_k)
\end{equation}
where $\Gamma$ is a phenomenological parameter describing the cavity photon leakage.
The dynamics of the density matrix operator of the whole quantum system
($H=H_0+H_{int}$) in which the damping of the excited states is taken into
account is described by means of the following standard master equation
\begin{equation}
\begin{split}
\frac{d}{dt} \rho (t)&= -i[H,\rho (t)]\\
& + \gamma_X\left(\frac{1}{2}\{|G\rangle,|X_R\rangle\}_{\rho(t)} + \frac{1}{2}\{|G\rangle,|X_L\rangle\}_{\rho(t)}\right) \\
& + \gamma_B\left(\frac{1}{2}\{|X_R\rangle,|B\rangle\}_{\rho(t)} + \frac{1}{2}\{|X_L\rangle,|B\rangle\}_{\rho(t)}\right) \\
& + \Gamma                  \{\mathbbm{1},\alpha_k^{\dagger}\}_{\rho(t)}
\end{split}
\label{master}
\end{equation}
where the operator \[\{u,v\}_f = 2uu^{\dagger}v^{\dagger}fv - fvv^{\dagger} -
vv^{\dagger}f\] and  $ \gamma_X$, $ \gamma_B$ are the damping constants from
the exciton and biexciton, respectively. The last term of eq.(\ref{master})
represents the photon leakage from the cavity to the environment.

We focus on the steady state solution of the master equation achieved, at
sufficiently large $t$, by vanishing the time derivative, of the coupled
non-linear differential equations of the master equation and practically
solving an algebraic system of non-linear equations.

\section{ENTANGLED PHOTON GENERATION}

We work in the weak coupling regime assuming all the coupling constants much
smaller than the characteristic energy, $\epsilon_0$ of the system under
investigation. Moreover, we consider the weak fields limit of very small
amplitudes of both the external electric fields. Both conditions should be met
as under these conditions eq.(\ref{master}) is a valid master equation. 
Now, we turn our attention to the emitted photon of the QD-cavity system.  As
we are in the weak field limit, it is sufficient to consider only the subspaces
of zero, one and two excitations of QD and cavity simultaneously. Actually it
is more convenient to work with the dressed states of the QD-cavity system and
assume only cascade transitions between neighboring subspaces. Hence the
corresponding transitions and emissions of a k-polarized photons are
represented by the following operators
\begin{equation}
(|n\rangle \langle n|) \alpha_k  (|m\rangle \langle m|) = |n\rangle (\langle n| \alpha_k  |m\rangle) \langle m| = \gamma_{n,m} |n\rangle \langle m|,
\end{equation}    
where $ \gamma_{n,m}$ is the transition amplitude from the $ |n \rangle$ to the $|m \rangle$ ($ \gamma_{n,m}$ are summarized in Ref.\cite{ajiki5}, from eq.(15) till eq.(19)).
As the energy splitting of the dressed states in the two upper subsets is
unequal we can take advantage of this and create entangled photons. The idea is
the following. We use input field frequencies $ \Omega_L$ or $ \Omega_R$ tuned
to the lower state of single excitation (of energy $ \epsilon_0$ -  $g$). Then
the photon pairs generated via the upper state of single excitation  (of energy
$ \epsilon_0$ +  $g$) would have energies which are different from $\Omega_L$
and $\Omega_R$. Then by performing spectral filtering we extract these photons.
Hence, one possible scheme would be to tune $\Omega_L$ to the $ \epsilon_0$ -
$g$ state resonantly driving the $|G \rangle \to |L_- \rangle$ transition, and
consider spectral filtering of the $ (\Omega_R + \Omega_L) - (\epsilon_0 + g)$
and  $ \epsilon_0$ +  $g$. In order to estimate any property of the photon
pairs generated via the cascade process we need the reduced density matrix
where the bases are in the following order $| L( \omega_1) \rangle | R(
\omega_2) \rangle$,  $| R( \omega_1) \rangle | L( \omega_2) \rangle$,   $| L(
\omega_1) \rangle | L( \omega_2) \rangle$ and  $| R( \omega_1) \rangle | R(
\omega_2) \rangle$, with photon frequencies $ \omega_1 = (\epsilon_0 -g)$ and
$ \omega_2 = \Omega_R + \Omega_L -(\epsilon_0 -g)$.

Hence our major task is to find the density matrix, $\rho_{\mathrm{ph-pairs}}$ of this
four dimensional ($| L \rangle | R \rangle$,  $| R \rangle | L \rangle$,   $| L
\rangle | L \rangle$, $|R\rangle|R\rangle$) Hilbert space, as once we know the
$\rho_{\mathrm{ph-pairs}}$ density matrix we can extract any information for the
classical and quantum correlations of the photon pairs generated in the
QD-cavity system. The $\rho_{\mathrm{ph-pairs}}$ density matrix can be calculated by
properly tracing out the $\rho$ density matrix describing the whole QD-cavity
system. As the photon pairs $|nm \rangle$ are generated through cascade
transitions through the $|R+ \rangle$ or $|L+ \rangle$ dressed states, we can
describe this process by transition operators, $T(n m)$ acting on the total
Hamiltonian, $H$. For example the transition operator which creates the $|RR
\rangle$ pair is given by
\begin{equation}
	\begin{split}
		T(RR) & = \gamma_{G;R+} \Ket{G} \Bra{R+} \sum_{k=-,+} \gamma_{R+;RRk} \Ket{R+} \langle RRk|   \\
& = \gamma_{G;R+} \sum_{k=-,+}  \gamma_{R+;RRk} |G \rangle \langle RRk|  \\
\end{split}
\end{equation}     

The  density matrix, $\rho_{\mathrm{ph-pairs}}$ can be represented as
\begin{equation}
	\rho_{\mathrm{ph-pairs}} = \sum_{n',m',n,m=R,L} (\langle |n' m' \rangle \langle nm| \rangle)   |n' m' \rangle \langle nm|
\end{equation}     
where the coefficients $\langle |n' m' \rangle \langle nm| \rangle$ are proportional to the  $\mathrm{Tr}[\rho\ T^+ (n'm') T(nm)]$. The proportionality coefficients are determined from the condition  $\mathrm{Tr}[\rho_{ph-pairs}]=1$.

\section{NUMERICAL RESULTS AND DISCUSSION}

\begin{figure}[htbp!]
\centering
\includegraphics[width=9 cm]{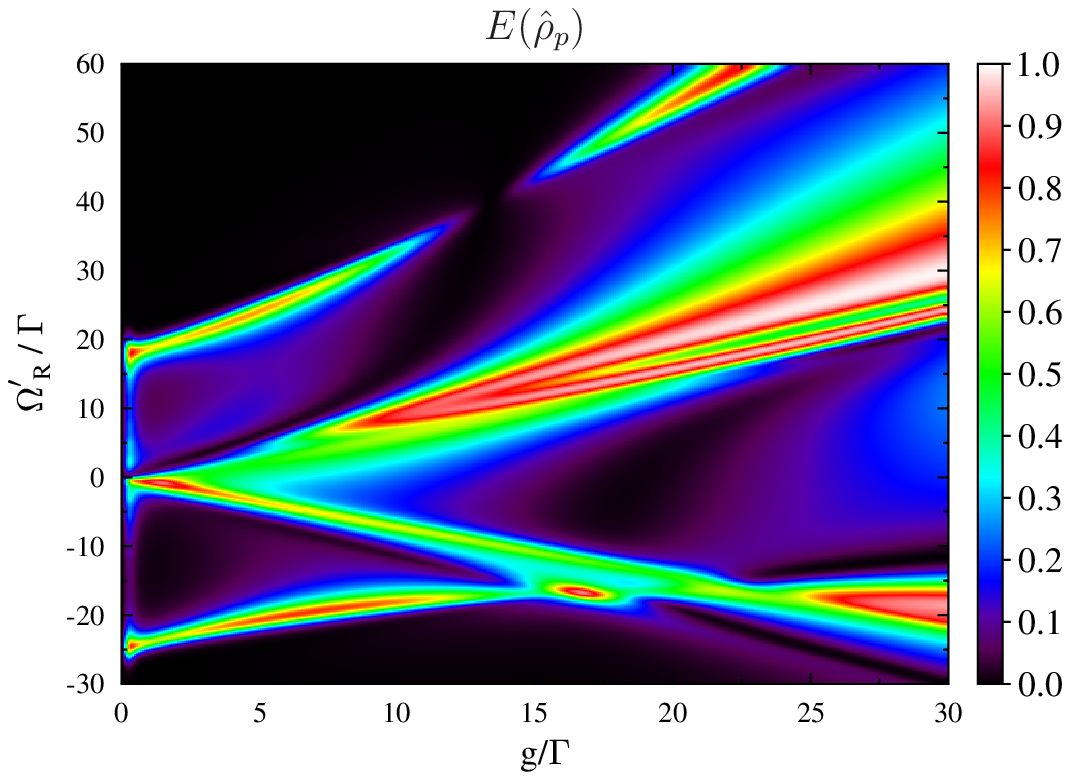} 
\includegraphics[width=9 cm]{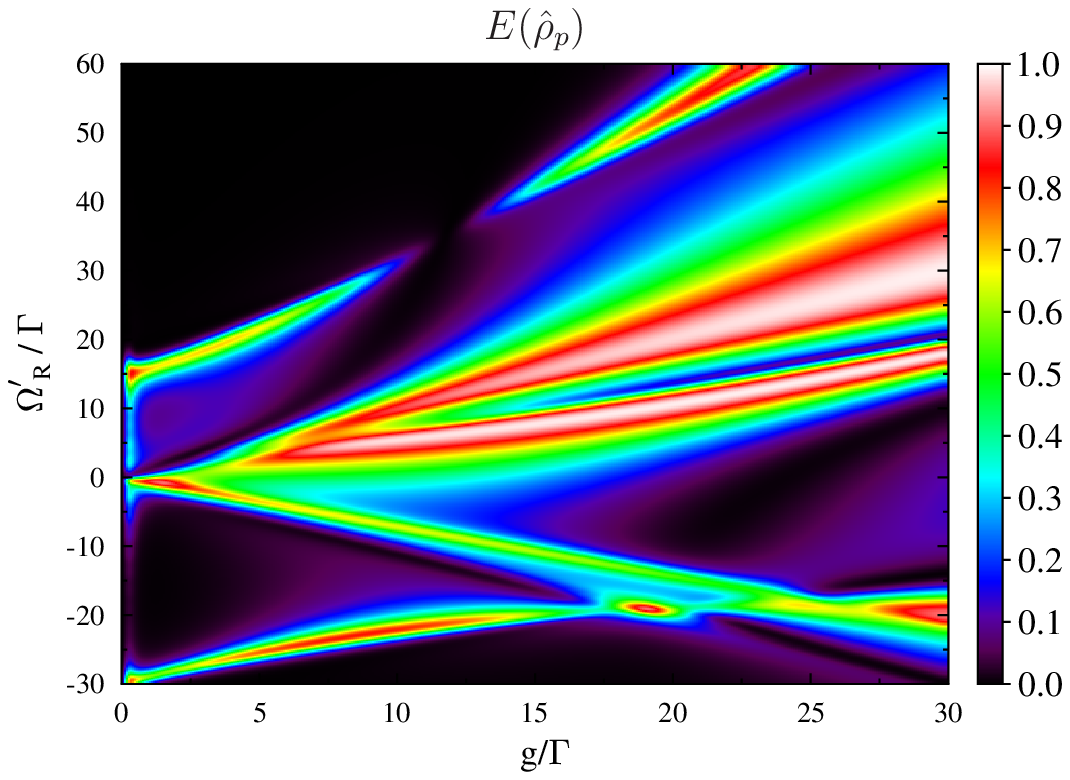} 
\includegraphics[width=9 cm]{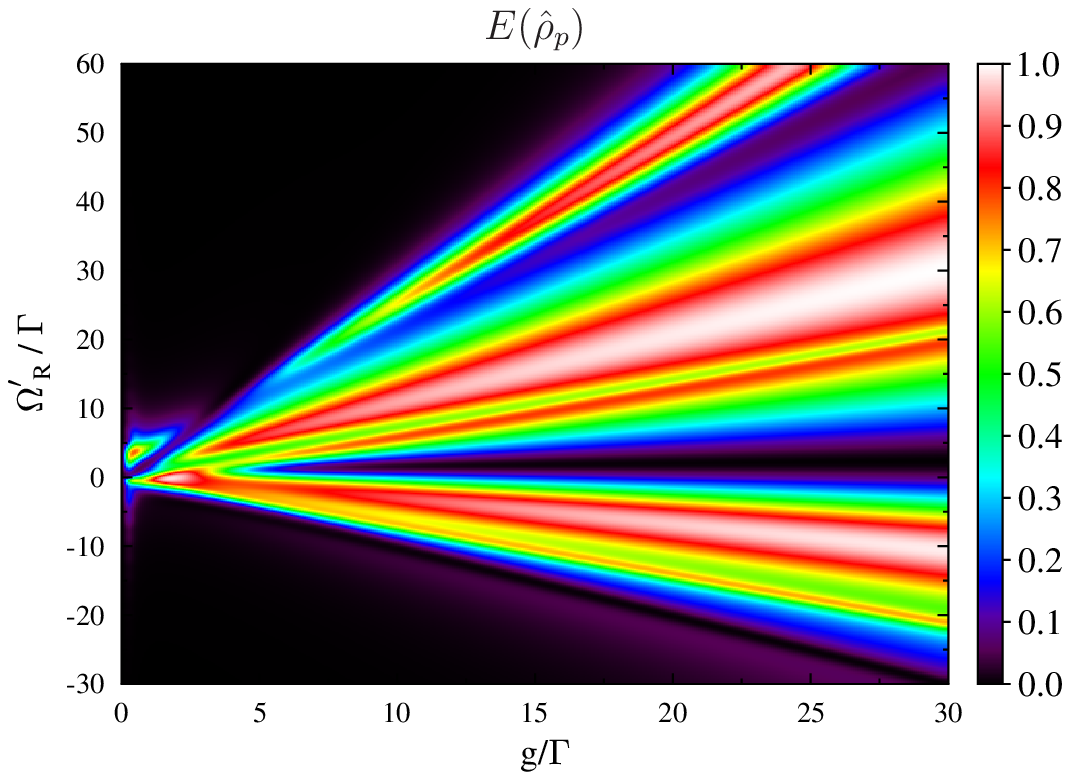} 
\caption{The degree of entanglement (EoF) for the QD-cavity system as a function
of the $g/\Gamma$ for (a) $\Delta_B=7\Gamma$,(b) $\Delta_B=15\Gamma$ and (c)
$\Delta_B=150\Gamma$. The highest degree of entanglement, for all cases, is
observed at the center of the diagram and more precisely for the dressed
states $|S\rangle$ and $|T_2\rangle$. }
\label{fig:Db}
\end{figure}

In this section, we report the degree of entanglement (EoF) of the photon pairs
generated from the QD-cavity system for different values of system parameters.
Figure.\ref{fig:Db} shows EoF as a function of $g$ and
$\Omega'_R=\Omega_R-\epsilon_0$ for fixed values of $g_B=15\Gamma$ and the
input field frequency at $\Omega_L=\epsilon_0-g$ (i.e. tuned at the lowest
possible transition) and with three different values of the biexciton binding
energy. A relatively low value of $\Delta_B$ of $7\Gamma$ (e.g. InAs QD) at
fig.\ref{fig:Db}(a), an intermediate value of $15\Gamma$ (e.g. GaAs/AlAs QD) at
fig.\ref{fig:Db}(b) and a high one of $150\Gamma$ (e.g. CuCl QD) at
fig.\ref{fig:Db}(c).

\begin{figure}[htbp!]
\centering
\includegraphics[width=9 cm]{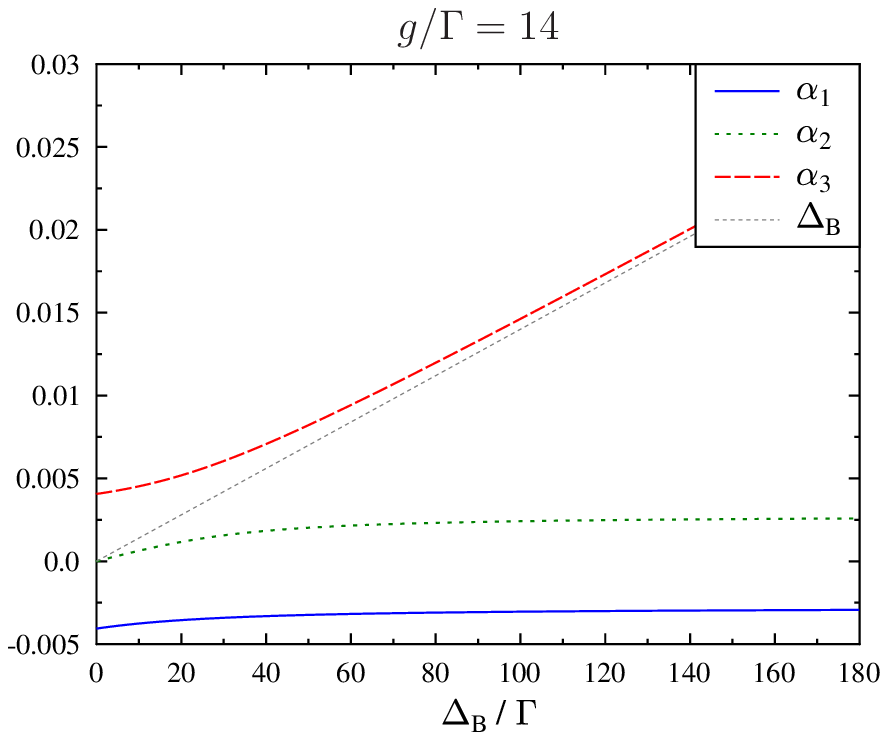}
\includegraphics[width=9 cm]{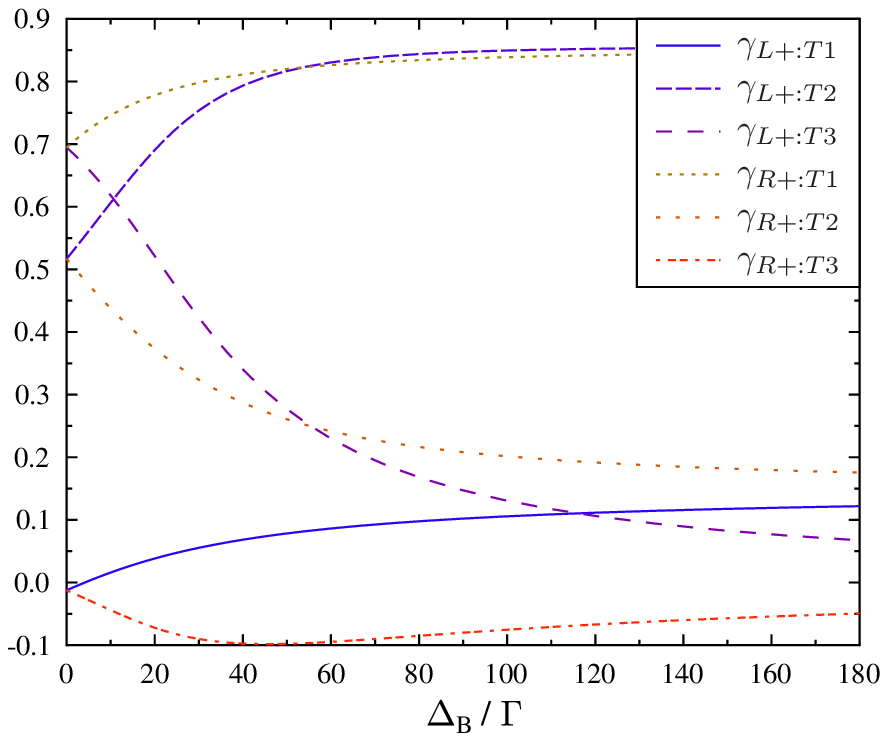}
\caption{(a) The energy of the dressed T-states and (b) the transition
amplitudes of the dressed T-states to $R+$ and $L+$ states as a function of the
biexciton binding energy.}
\label{fig:ai}
\end{figure}

 In all figures ((a) to (c)) it is apparent that the largest values of the
 entanglement reported by EoF occur along the lines which correspond to
 resonant transitions of the upper four dressed states to the intermediate
 state (of energy $\epsilon_0+g$). The four major lines denote resonant
 excitations for the dressed state $|T_1\rangle$ (top curve with positive
 slope), $|S\rangle$ (second curve), $|T_2\rangle$ (third curve) and
 $|T_3\rangle$ (bottom curve with positive slope). There are two more lines
 denote resonant excitations for the dressed state $|RR_-\rangle$ (curve with
 negative slope between the intermediate single and triple curves)
 and$|RR_+\rangle$ (bottom curve with negative slope in plots (a) and (b)). For
 the high value of the binding energy (plot (c)) the two latter curves
 disappear or one changes slope. In the low and intermediate binding energies
 (plots (a) and (b)), we find that there is a gap where the EoF becomes zero.
 This disappearance of the entanglement originates by the fact that the
 transition amplitude $\gamma_{L+;T1}$ becomes zero. Under this condition we
 obtain that for $g=g_-$, with
 $g_-=\frac{1}{4}(\sqrt{\Delta_B^2+16g_B^2}-\Delta_B)$ \cite{ajiki5}. As
 expected, when we increased the binding energy this gap moved towards to the
 beginning of the excitation line of $|T_1\rangle$. Moreover, for small values
 of $\Delta_B$ the dressed states $|T_2\rangle$ and $|S\rangle$ are almost
 degenerated but in our case this is not true. Indeed, there are two regions of
 maximum entanglement at the center of the diagram which means that the two
 dressed states are not degenerated. This happens, as from \cite{ajiki5} we
 know the relation $a_1<0<a_2<\Delta_B<a_3$. As a result, for small values of
 the binding energy $\Delta_B$, the value of $a_2$ is almost zero, thus the
 dressed states $|S\rangle$ and $|T_2\rangle$ are almost degenerate, which is
 not happen in our case as mentioned above. 

By systematically investigating cases of binding energy higher than the one of
$150\Gamma$ (figure\ref{fig:Db}(c)), we have found that the EoF diagram
remains practically unchanged. The explanation of the latter behaviour is
given by means of figure \ref{fig:ai} where we reveal that the dressed states
energies (fig. \ref{fig:ai}(a)) and the transition rates (fig. \ref{fig:ai}(b))
remain unchanged above an binding energy at $160\Gamma$.
 
\begin{figure}[htbp!]
\centering
\includegraphics[width=9 cm]{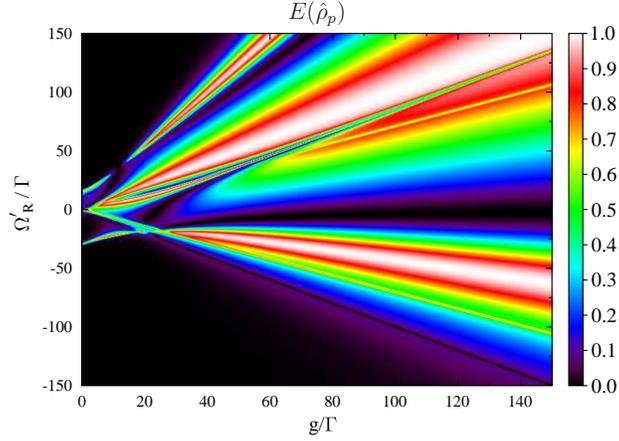}
\caption{The degree of entanglement (EoF) for the QD-cavity system as a function
of the $g/$$\Gamma$ for $\Delta_B=15\Gamma$, $g_B=15\Gamma$,
$\gamma_X = \gamma_B = 0.1\Gamma$, $E_R = E_L = 0.02  \sqrt{\Gamma}$. }
\label{fig:canvas}
\end{figure}

Figure \ref{fig:canvas} shows the entanglement of formation (EoF) for a wider
range of g and $\Omega'_R$. Now we have also set $\gamma_X=\gamma_B=0.1\Gamma$.
A thin line of maximum EoF surrounded by a region where the entanglement is
zero or very small, is observed at the center of the diagram between of
excitation lines of $|S\rangle$ and $|T_2\rangle$. Additionally, we observe
other thin lines of zero entanglement just under of the above line at the
center of the diagram too, as well as under the line of excitation with the
dressed state $|T_3\rangle$. Actually under the line of $|T_3\rangle$ there are
two thin lines of zero entanglement. From
\cite{ajiki5} we know the relations $a_1<0<a_2<\Delta_B<a_3$ as well as
$a_1<-\sqrt{2}g<a_2<\sqrt{2}g<a_3$, where the $a_j$ is the difference between
$2\omega_0$ and the corresponding eigen-energies. By adding these relations we
also have $a_1<-g/\sqrt{2}<a_2<g/\sqrt{2}+\Delta_B/2<a_3$. These relations
indicate the possible values but also give information about the forbidden
values of $a_j$. More specifically, it is evident that $a_j$ cannot be zero for
example as must be greater or smaller than this value. As a consequence, there
are forbidden areas in this figure where the entanglement becomes zero. In one
of these lines, as mentioned above, there is a thin line of maximum. This
happens because for $\Omega'_R=g-\Delta_B$ there is a forbidden area  but also
there is resonance with the dressed state $|T_2\rangle$ with $\Omega'_R=g-a_2$,
with $0<a_2<\Delta_B$.

\begin{figure}[htbp!]
\centering
\includegraphics[width=9 cm]{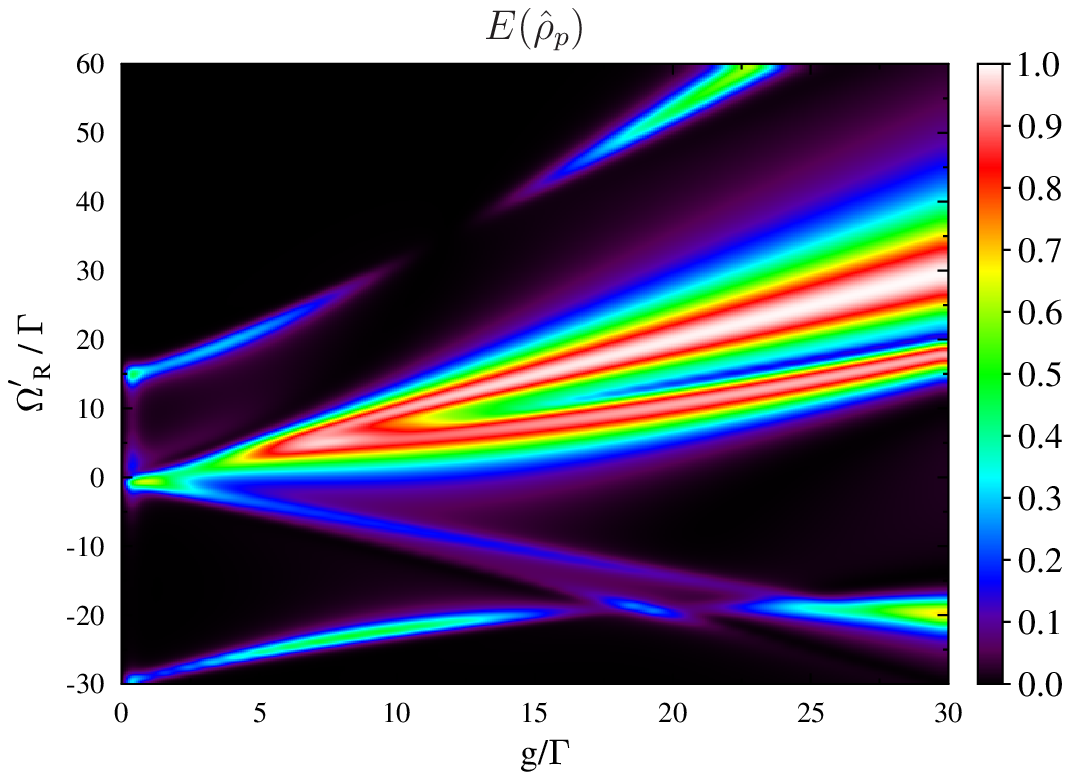}
\includegraphics[width=9 cm]{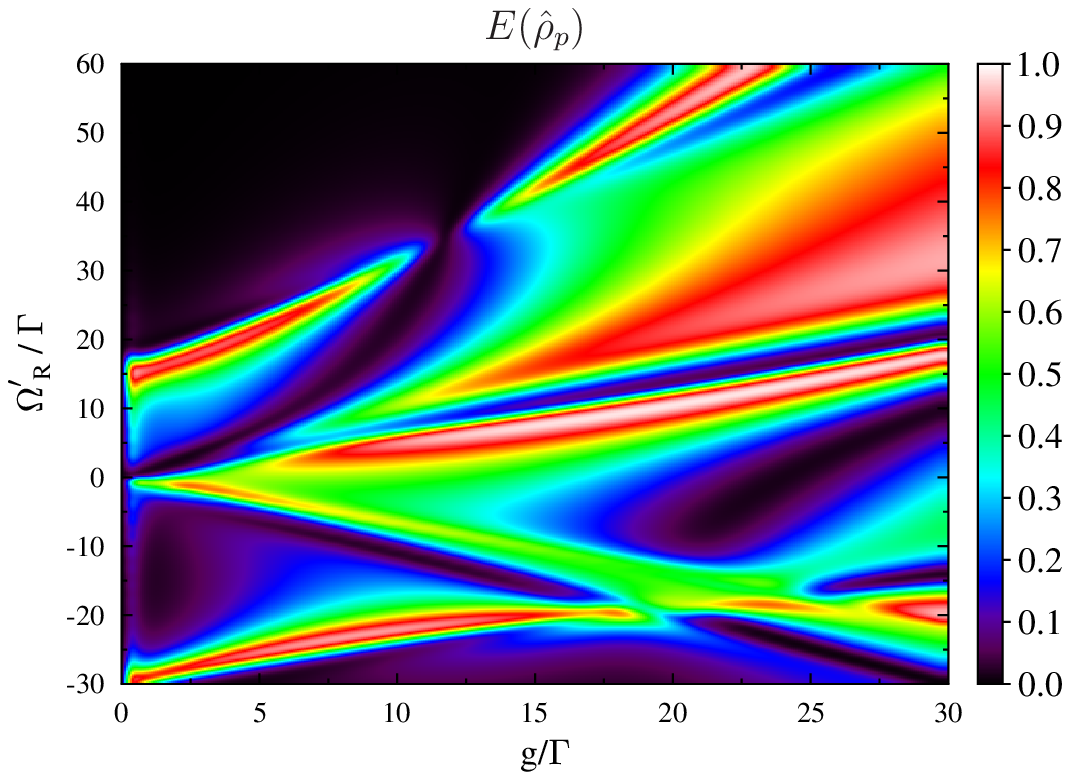}
\caption{Entanglement of formation (EoF) for (a) $E_R=0.01\sqrt{\Gamma}$ and
$E_L=0.02\sqrt{\Gamma}$. There are maximums only for resonant with $|S\rangle$
and $|T_2\rangle$ and for (b) $E_L=0.01\sqrt{\Gamma}$ and
$E_R=0.02\sqrt{\Gamma}$. The entanglement has completely different behaviour in
comparison to the previous case (a).Although should be noted that in both cases
only two Bell-states, $|B_1\rangle=\frac{1}{\sqrt{2}}(|RL\rangle+|LR\rangle)$
and $|B_2\rangle=\frac{1}{\sqrt{2}}(|RL\rangle-|LR\rangle)$ are generated.}
\label{fig:EREL}
\end{figure}

Finally, we systematically investigate the effect of the laser intensities.
Figure \ref{fig:EREL}(a) shows the degree of entanglement as a function of g
and $\Omega'_R$ with $E_R=0.01\sqrt{\Gamma}$ and $E_L=0.02\sqrt{\Gamma}$
whereas in figure \ref{fig:EREL}(b) is $E_R=0.02\sqrt{\Gamma}$ and
$E_L=0.01\sqrt{\Gamma}$ and all the other parameters are exactly the same. As
it is easily observable the two diagrams are not symmetric to each other. This
is because the left-polarized laser field is tuned to the state $|L_-\rangle$
as we have $\Omega_L=\omega_0-g$, thus we have more excitations and as a
consequence more generated entangled photon pairs in the second case. For this
reason, in figure \ref{fig:EREL}(b) we have many regions with high degree of
entanglement. We observe maximum EoF along the lines which indicate resonant
with the dressed states $|S\rangle$, $|T_j\rangle$ (with j=1,2,3) as well as
the co-polarized state $|RR_-\rangle$.

It is worth noting that the gap where the EoF becomes zero across the line of
$|T_1\rangle$ still remains but this is not true for the maximum along the line
of $|RR_-\rangle$ where instead of a maximum we observe a minimum. This is
because the pairs which generated from the state $|RR_-\rangle$ or
$|LL_-\rangle$ are not entangled any more since we use lasers with different
amplitudes thus, the probability to have excitation with a right-polarized
photon is greater than the corresponding probability of an excitation with a
left-polarized photon. Additionally there is a thin line of zero entanglement
between the excitation lines of $|S\rangle$ and $|T_2\rangle$. On the other
hand, in figure \ref{fig:EREL} there are only two maximums along the excitation
lines of $|S\rangle$ and $|T_2\rangle$. Obviously, the results will be reversed
if one tunes $\Omega_L$ laser frequency to the $ \epsilon_0$ +  $g$ state
resonantly driving the $|G \rangle \to |L_+ \rangle$ transition, and consider
spectral filtering of the $(\Omega_R + \Omega_L) - (\epsilon_0 + g)$ and  $
\epsilon_0$ -  $g$.

\section{Conclusions}

In summary, in the present work, we theoretically investigated  the generation of polarization-entangled photons from a microcavity in which a QD is embedded in, radiated by two laser fields of opposite circular polarization (left/right). Our theoretical model relies on the dressed states scheme of the QD nanostructure considered as a four-level system and containing energies up to the order of the two exciton states). 
 By systematically investigate the entanglement of formation, for various values of the important parameters of the quantum dot system as the bi-exciton binding energy and the intensities of the applied fields we have shown that there is a strong dependence of the quantum correlations on them.

\section{acknowledgements}
We would like to thank Petros Androvitsaneas and Charis Anastopoulos for helpful discussions and suggestions during the production of the present work.

{}

\end{document}